# COMPACT: Concurrent or Ordered Matrix-based Packing Arrangement Computation Technique

Gokhan Serhat [*]

Max Planck Institute for Intelligent Systems, Heisenbergstr. 3, 70569 Stuttgart, Germany

**Abstract**

Packing optimization is a prevalent problem that necessitates robust and efficient algorithms that are also simple to implement. One group of approaches is the raster methods, which rely on approximating the objects with pixelated representations. Although they are versatile in treating irregular geometries, the raster methods received limited attention in solving problems involving rotatable objects, where available studies generally analyze only right-angled rotations. In addition, raster approximation allows the use of unique performance metrics and indirect consideration of constraints, which have not been exploited in the literature. This study presents the new Concurrent or Ordered Matrix-based Packing Arrangement Computation Technique (COMPACT). The method relies on raster representations of the objects that can be rotated by arbitrary angles, unlike the right-angled rotation restrictions imposed in many existing packing optimization studies based on raster methods. The raster approximations are obtained through loop-free operations that improve efficiency. Besides, a novel performance metric is introduced, which favors efficient filling of the available space by maximizing the internal contact between the objects as well as the contact between the objects and domain boundaries. Moreover, the objective functions are exploited to discard overlap and overflow constraints and enable the use of unconstrained optimization methods. Several test problems involving concurrent and ordered packing of multiple rectangular and circular objects into square bins are investigated. The results show that the proposed technique performs effectively in determining the packing arrangements.

1. **Introduction**

Packing optimization is an important problem class that concerns various disciplines including warehouse management, shipping, and design of engineering products. The widespread nature of this problem requires robust and efficient algorithms that are easy to utilize.

Various techniques have been developed to solve different packing optimization problems. A large amount of available studies addresses the problem of packing two-dimensional rectangular objects into rectangular bins while keeping the object edges orthogonal to the domain boundaries. For instance, Chen and Huang (2007) computed solutions for two-dimensional rectangle packing problems by introducing a search algorithm that places the rectangles into the container one by one using a corner occupation strategy. Later, Leung, Zhang, Zhou, and Wu (2012) proposed a hybrid heuristic algorithm that combines greedy strategy and simulated annealing methods to solve the rectangular knapsack packing problem. Recently, Chen et al. (2019) developed a deterministic heuristic algorithm for solving the two-dimensional rectangular packing problems that require maximizing the filled portion of the rectangular domain.

Packing problems concerning nonrectangular objects and domains were also prevalently investigated. For example, Del Valle, Queiroz, Miyazawa, and Xavier (2012) developed heuristic approaches for solving 2D cutting/packing problems involving irregular objects. Later, Chen et al. (2018) proposed a heuristic algorithm for solving circle packing problems for placing multiple identical circles into the smallest circular container. Lately, Bouzid and Salhi (2020) investigated orthogonal packing of rectangular objects into a circular space of fixed radius.

---

[*] Email address: serhat@is.mpg.de (Gokhan Serhat)

The complexity of the packing problems notably increases when the nonorthogonal rotation of the objects is allowed in addition to the translation (Cherri et al., 2016). These problems received relatively lesser attention compared to the problems involving fixed object orientations or orthogonal rotations. In one study, Zhang and Zhang (2009) introduced a 2D packing optimization technique called finite-circle method, which approximates polygons with a set of circles and converts the non-overlapping constraints into simple constraints between circles. Martins and Tsuzuki (2010) used a heuristic approach to minimize the waste space during the rotational placement of two-dimensional polygons into irregular containers. In another study by Martinez-Sykora et al. (2017), a strategy was proposed to solve bin packing problems with irregular pieces that can be freely rotated.

The difficulties in the formulations for packing problems involving irregular objects have led to the development of raster point methods, which rely on approximating the objects on a discretized grid. With these methods, packing configurations can be represented by assigning zeroes to the free domain cells and different values to the cells where a single object is present or multiple objects overlap. The key advantage of raster techniques is their applicability to the problems involving objects with complex shapes, where the main drawback is their resolution-dependent precision (Leao et al., 2019). In the literature, the research on packing optimization with raster methods is limited compared to the available work based on other techniques. In a particular study, Toledo et al. (2013) proposed a discrete modeling approach called the Dotted-Board Model, which relies on representing the bin with a two-dimensional grid and using the plausible grid points for the placement of arbitrary polygons. Later, Mundim, Andretta, and de Queiroz (2017) extended the Dotted-Board Model for packing non-polygonal objects with curved edges and used it within a heuristic approach based on genetic algorithms. Recently, Sato, Martins, Gomes, and Tsuzuki (2019) developed the raster overlap minimization algorithm to solve the irregular strip packing problem using a raster method that limits the solution space.

This study presents the new Concurrent or Ordered Matrix-based Packing Arrangement Computation Technique (COMPACT). In this framework, the object shapes are approximated by matrix-based (raster) representations. The objects can be rotated by angles that are not necessarily restricted to be right-angle multiples, unlike the conditions imposed in many existing packing optimization studies relying on raster methods. The raster approximations are obtained through loop-free operations that improve efficiency. Besides, a novel performance metric is introduced, which aids efficient filling of the available space by maximizing the internal contact between the objects as well as the contact between the objects and domain boundaries. Moreover, the possibility of exploiting the objective functions for discarding overlap and overflow constraints is explored to enable the use of unconstrained optimization methods. As case studies, the developed technique is utilized to pack multiple rectangular and circular objects in square domains using concurrent and ordered placement strategies. The results show that the proposed technique performs effectively in determining the packing arrangements. The organization of the study is as follows: The matrix-based object representation that forms the backbone of the developed technique is described in the following section. Section 3 covers the details regarding the formulation of the optimization problem. Section 4 covers the results from various case studies. In Section 5, the conclusions are stated and the directions for future studies are outlined.

## 2. Matrix-based object representation

The proposed technique relies on the raster representation of the objects in the matrix space. Let a rectangular design space be described by an empty matrix with $H$ rows of height and $W$ columns of width. A rectangular object having a height of $h$ rows and a width of $w$ columns can now be modeled within the domain by filling an area consisting of $h \times w$ entities with ones. The geometries should be approximated for curved shapes and polygons whose edges are not aligned with the rows and columns. Fig. 1 shows (a) exemplary matrix space approximations of a circle and a rectangle whose edges are aligned with the horizontal and vertical domain axes, and (b) the sparsity pattern of the matrix containing the objects.

The proposed technique does not restrict the edges of the rectangular objects to be aligned with the domain boundaries. Raster approximation of the rotated objects is exemplified with a $h = 0.4H = 0.4W$ by $w = 1.5h$ rectangle rotated by an angle of $\theta = 30°$ around its top-left corner. Fig. 2 shows the sparsity patterns of this object for domain sizes of (a) 20 × 20, (b) 50 × 50, and (c) 100 × 100, which results in 92, 595, and 2414 nonzero entities ($N_{nz}$), respectively. As seen in this figure, the shape of the rotated geometry is more accurately captured with increasing resolution. A mathematical convergence can also be observed by calculating the ratio of nonzero entities to the total number of domain entities. When an unrotated rectangle is considered, the exact filling ratio (the proportion of ones to the total number of elements) can be calculated as $hw/HW = 0.24$. For the presented three cases, the filling ratios are computed as 92/400 = 0.2300, 595/2500=0.2380, and 2414/10000 = 0.2414, which converge to the exact value with higher resolution. However, increasing the number of raster points inherently lengthens the computation durations (Leao et al., 2019). Therefore, excessive discretization resolutions should be avoided to keep the computation durations within reasonable limits.

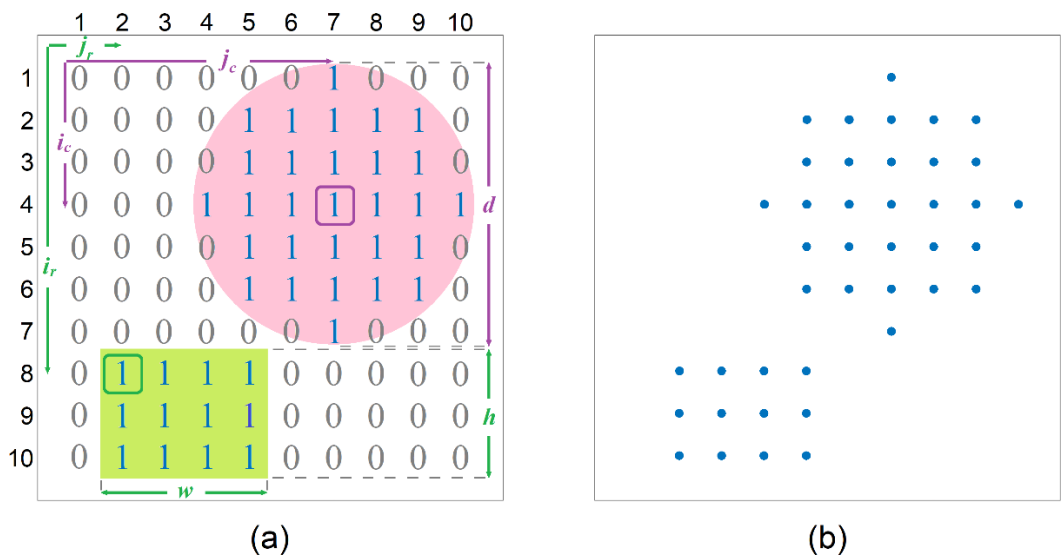

**Fig. 1.** Matrix space approximations of a rectangle with $h = 3$ rows height and $w = 4$ columns width, and a circle with $d = 7$ cells diameter. The upper-left corner of the rectangle and center of the circle are located on ($i_r = 8$, $j_r = 2$) and ($i_c = 4$, $j_c = 7$), respectively. (b) The sparsity pattern of the matrix containing the objects.

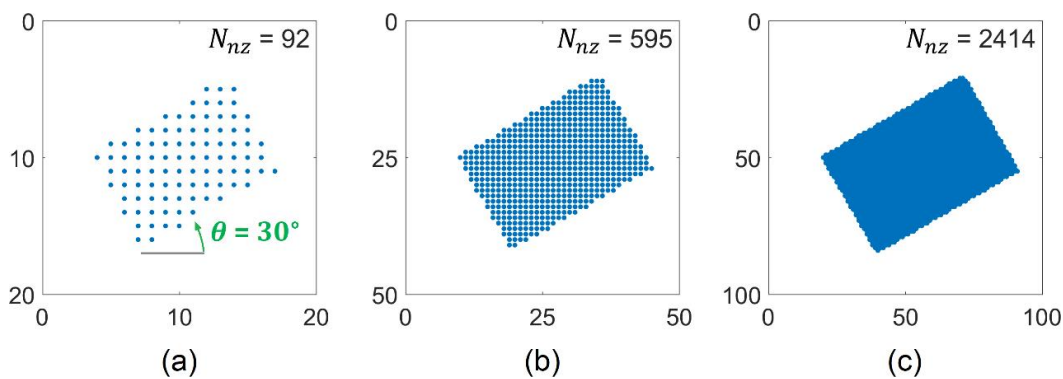

**Fig. 2.** Sparsity patterns of a rectangle rotated by an angle of $\theta = 30°$ around its top-left corner for domain sizes of (a) 20 × 20, (b) 50 × 50, and (c) 100 × 100; which result in 92, 595, and 2414 nonzero entities, respectively. The rectangles have a size of ($h = 0.4H$, $w = 1.5h$).

The raster approximations for the rectangles are attained by performing vector inequality operations and elementwise matrix multiplications. In the present study, only top-left corner coordinates of the rectangles are restricted to be integers. The other corners do not necessarily coincide with the matrix grid unlike the conditions enforced with certain raster-based methods such as the Dotted-Board Model (Toledo et al., 2013; Mundim, Andretta, and de Queiroz, 2017).

For a rectangle rotated counterclockwise by $\theta$ degrees, the row and column coordinates of the bottom right corner ($i_\theta$ and $j_\theta$) can be found as

$$i_\theta = i_r + h\cos(\theta) - w\sin(\theta) \\ j_\theta = j_r + h\sin(\theta) + w\cos(\theta) \tag{1}$$

Let $r_H$ be a row vector and $c_W$ be a column vector with elements ranging from 1 to $H$ and 1 to $W$, respectively. Then, the lower and upper boundaries for the rows ($i_l$ and $i_u$) as well as the columns ($j_l$ and $j_u$) can be computed as

$$i_l = \lfloor i_r - \tan(\theta)(c_W - j_r) \rceil \\ i_u = \lfloor i_\theta - \tan(\theta)(c_W - j_\theta) \rceil \\ j_l = \lfloor j_r + \tan(\theta)(r_H - i_r) \rceil \\ j_u = \lfloor j_\theta + \tan(\theta)(r_H - i_\theta) \rceil \tag{2}$$

where the brackets '$\lfloor\ \rceil$' signify rounding towards the nearest integer. Note that, $i_l$ and $i_u$ are column vectors where $j_l$ and $j_u$ are row vectors. Using the calculated boundary vectors, four constituent matrices can be obtained through the following vector inequalities:

$$D_1 = (r_H \geq i_l),\ D_2 = (r_H < i_u), \\ D_3 = (c_W \geq j_l),\ D_4 = (c_W < j_u) \tag{3}$$

The final object representation is attained by performing element-by-element multiplication (denoted by operator '∘') of calculated constituent matrices:

$$D = D_1 \circ D_2 \circ D_3 \circ D_4 \tag{4}$$

The described approach avoids using loops, which in turn substantially accelerates the computations. Fig. 3 demonstrates the constituent matrices and final representation for a sample rectangular object.

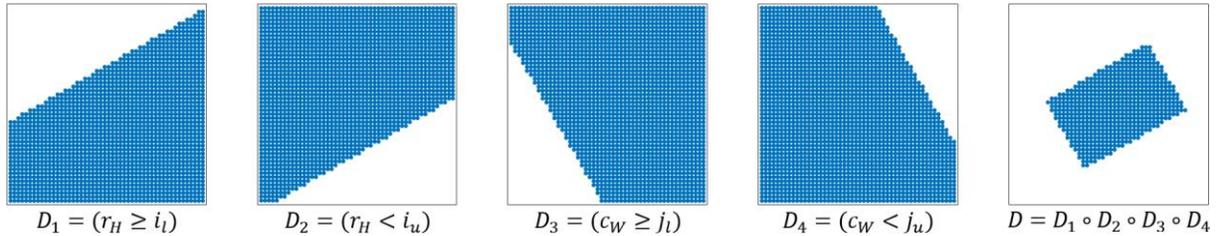

**Fig. 3.** The constituent matrices ($D_1, D_2, D_3, D_4$) and resulting representation ($D$) for an exemplary rectangle. The final object representation is attained by performing elementwise multiplication of four constituent matrices obtained by using simple vector inequalities.

For circles, the raster approximations can be directly obtained without the necessity of calculating multiple constituent matrices. Let $r_1$ be a row vector of ones with a length of $H$, $c_1$ be a column vector of ones with a length of *W*, and $I_1$ be an *H* by *W* matrix of ones. Then, the distance matrices along rows ($s_i$) and columns ($s_j$), and resultant distance matrix (*s*) can be calculated as follows:

$$s_i = i_c I_1 - r_H c_1, s_j = j_c I_1 - r_1 c_W, \qquad s = \sqrt{s_i \circ s_i + s_j \circ s_j} \qquad (5)$$

Finally, the matrix entities lying inside the circle can be filled with ones using the relation:

$$D(s \leq d/2) = 1 \qquad (6)$$

## 3. Optimization problem

*3.1. Objective functions*

*3.1.1. Total ones (T1)*

The first objective function is simply defined as the number ones in the domain matrix ($N_1$):

$$f_{T1} = N_1 \qquad (7)$$

The maximization of this metric is equivalent to occupying the design space as much as possible without causing overlaps. However, this technique cannot be used in sequential placement as it would rank all feasible locations identically and create an ill-posed problem.

*3.1.2. Distance from the center (DC)*

As the second objective, the rectilinear (Manhattan) distance between the centers of the objects and domain is maximized. This approach favors filling the areas next to the boundaries starting from the corners. Let the row and column indices of the domain center be $i_D^o$ and $j_D^o$, respectively. For a rectangle, the central row ($i_r^o$) and column ($j_r^o$) indices can be calculated as

$$\begin{aligned} i_r^o &= i_r + h/2 \cos(\theta) - w/2 \sin(\theta) \\ j_r^o &= j_r + h/2 \sin(\theta) + w/2 \cos(\theta) \end{aligned} \qquad (8)$$

Using the central indices, the distance function to be minimized is defined as

$$f_{DC}^r = |i_r^o - i_D^o| + |j_r^o - j_D^o| \qquad (9)$$

For a circular object, the distance function can be written as follows:

$$f_{DC}^c = |i_c - i_D^o| + |j_c - j_D^o| \qquad (10)$$

*3.1.3. Distance to bottom-left (DBL)*

The third metric is defined as the rectilinear distance between the objects and the bottom-left corner of the domain. Minimization of this distance has been prevalently used in the previous packing studies (Jakobs, 1996; Hopper and Turton, 1999). The negative of the distance is used to be consistent with the other utilized metrics, which are all maximized in the optimization. Therefore, the objective functions for the rectangles ($f_{DBL}^r$) and circles ($f_{DBL}^c$) can be expressed as follows:

$$\begin{aligned} f_{DBL}^r &= i_r - j_r \\ f_{DBL}^c &= i_c - j_c \end{aligned} \qquad (11)$$

This objective is only used with sequential packing.

### 3.1.4. Adjacent ones (A1)

The number of adjacent ones is introduced as the last objective function, which correlates with the overall contact within the domain. This metric promotes efficient utilization of the available space and it is formulated by taking advantage of the raster representation. The objective function to be maximized ($f_{A1}$) is simply defined as the sum of the entities of matrix $F$, which is the elementwise multiplication of four submatrices of the current domain matrix $D$:

$$F = D(1:H-1, 1:W-1) \circ D(2:H, 1:W-1) \circ D(1:H-1, 2:W) \circ D(2:H, 2:W)$$

$$f_{A1} = \sum_{i=1}^{H-1} \sum_{j=1}^{W-1} F_{ij} \tag{12}$$

In Fig. 4, the principle behind the developed contact maximization technique demonstrated for two example domains containing rectangles (a) in contact and (b) separated by zeroes.

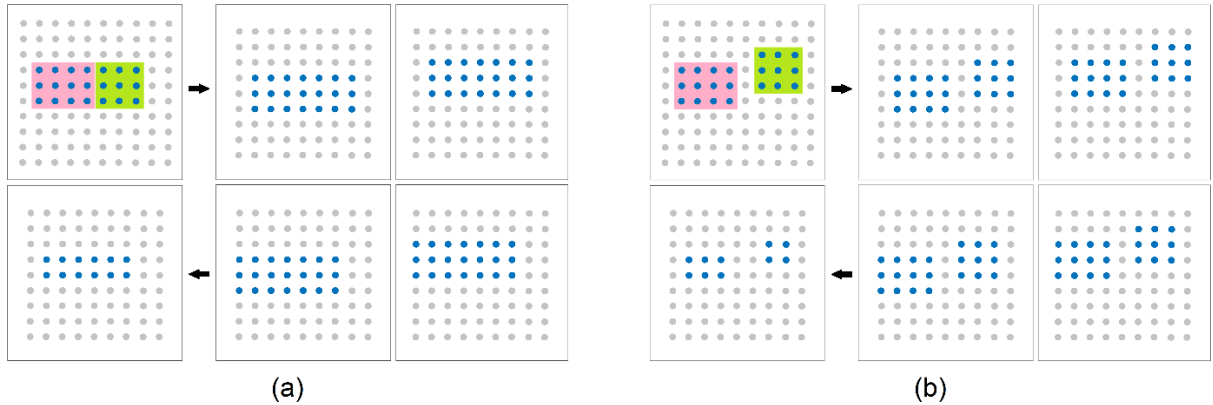

(a)          (b)

**Fig. 4.** Demonstration of the principle behind maximizing adjacent ones using two sample domains that contain rectangles (a) in contact (b) separated by zeroes. The sparsity patterns for domain matrices, four submatrices, and elementwise multiplications of the submatrices are shown at the top-left corners, right sides, and bottom-left corners of the subfigures, respectively. The contacting rectangles produce 12 nonzero entities while the separated ones yield 10 nonzero entities.

The proposed adjacent ones metric provides several advantages:
- It outperforms other metrics in solving certain problems,
- It is versatile delivering god results for several different cases,
- It is applicable for both concurrent and ordered optimization strategies,
- It can be used with or without explicit constraints while performing well in many unconstrained cases.

### 3.2. Design variables

The design variables considered within the optimization are the row and column indices for the upper-left corners of the rectangles ($i_r$'s and $j_r$'s) and centers of the circles ($i_c$'s and $j_c$'s). When the rotation is allowed for the rectangles, the angles of rotation ($\theta_r$'s) are considered as additional design variables. For $R$ rectangular and $C$ circular objects, all the design variables can be assembled into a single vector $x$ as follows:

$$x = \begin{bmatrix} i_r^1 & j_r^1 & \theta_r^1 & \cdots & i_r^R & j_r^R & \theta_r^R & i_c^1 & j_c^1 & \cdots & i_c^C & j_c^C \end{bmatrix}^{\mathrm{T}} \tag{13}$$

*3.3. Bounds on the design variables*

The domain of design variables is defined through lower and upper bounds. For rectangular objects, the values of the upper left corner coordinates and the rotation angles are bounded as follows:

$$\begin{aligned} 1 \leq i_r \leq H \\ 1 \leq j_r \leq W \\ -\pi/2 \leq \theta \leq \pi/2 \end{aligned} \quad (14)$$

For circular objects, the bounds on the central indices are defined as

$$\begin{aligned} \lfloor d/2 \rfloor \leq i_c \leq H - \lfloor d/2 \rfloor \\ \lfloor d/2 \rfloor \leq j_c \leq W - \lfloor d/2 \rfloor \end{aligned} \quad (15)$$

*3.4. Nonlinear constraints*

Two sets of nonlinear constraints are considered while solving the problem. The first set ($g_1$) ensures that the objects are contained within the design space, and the second set ($g_2$) enforces avoiding overlaps. These two sets are stacked together within a vector $g$ that should be smaller than zero:

$$g(x) = \begin{bmatrix} g_1(x) \\ g_2(x) \end{bmatrix} \leq 0 \quad (16)$$

*3.4.1. Containment assurance*

When the rotational motion is permitted for the rectangular objects, the lower and upper bounds of the position variables do not guarantee the prevention of overflowing. Therefore, a nonlinear constraint set is defined as

$$g_1 = \begin{bmatrix} -i + w\sin(\theta) + 1 \\ i + h\cos(\theta) - H \\ -j + 1 \\ j + w/\cos(\theta) + (h - w\tan(\theta))\sin(\theta) - W \end{bmatrix} \quad (17)$$

For circular objects, the containment is directly ensured through the limits imposed on the position variables.

*3.4.2. Overlap avoidance*

While placing an object into the domain, the matrices representing the new object and the domain with previous objects are simply summed together. An overlap is detected for the matrix elements with values higher than 1. Therefore, for $O$ total objects, the constraint for the overlap avoidance can be written as

$$g_2 = \sum_{k=2}^{O} N_k \quad (18)$$

While maximizing total ones or adjacent ones as the performance metrics, explicit collision constraints may not be necessary since the improvement in the objective function and reduction in the overlaps are correlated in certain problems.

*3.5. Optimizer*

The described optimization problem is solved using a genetic algorithm capable of handling the non-convex design space. Specifically, the 'ga' function of MATLAB's Optimization Toolbox is used. The genetic algorithms have previously been shown to be effective for escaping locally optimal design points arising in packing problems (Jakobs, 1996; Hopper and Turton, 1999). Since the optimal positions should correspond to the matrix indices, the design variables are constrained to be integers. For the cases involving rotation, the integer angle variable multiplied by the predefined angular increment controls the orientation.

## 4. Case studies

This section covers the results from different case studies that involve the packing of rectangular and circular objects into 100 × 100 square bins. Both concurrent and sequential optimization strategies are explored. In the sequential placement, the objects are ordered with the decreasing diagonal length or area. The total number of ones, objects' rectilinear distances from the center, objects' distance from the bottom-left corner, and the number of adjacent ones are used as the objective functions. The combined influence of the optimization settings on COMPACT's performance in solving different problems is investigated.

*4.1. Non-rotatable objects*

The developed methodology is initially utilized to obtain packing configurations for non-rotatable objects. The first study involves sets of rectangular objects whose heights and widths are given in Table 1. These sets are obtained by splitting one rectangle into two new ones at each step as demonstrated in Fig. 5. Hence, the number of rectangles is regularly increased from 4 to 9 while going from the first case (Fig. 5a) to the last one (Fig. 5f).

**Table 1** Heights and widths of the rectangles used in the sets used in the first study with restrained rotations.

| Number of objects | (Height, width) | | | | | | | | |
|---|---|---|---|---|---|---|---|---|---|
| 4 | (50, 60) | (50, 40) | (30, 100) | (20, 100) | | | | | |
| 5 | (50, 60) | (50, 40) | (30, 70) | (30, 30) | (20, 100) | | | | |
| 6 | (50, 60) | (50, 40) | (20, 70) | (10, 70) | (30, 30) | (20, 100) | | | |
| 7 | (50, 60) | (50, 40) | (20, 70) | (10, 70) | (30, 20) | (30, 10) | (20, 100) | | |
| 8 | (50, 60) | (50, 40) | (20, 70) | (10, 70) | (30, 20) | (30, 10) | (15, 100) | (5, 100) | |
| 9 | (50, 60) | (50, 40) | (15, 70) | (5, 70) | (10, 70) | (30, 20) | (30, 10) | (15, 100) | (5, 100) |

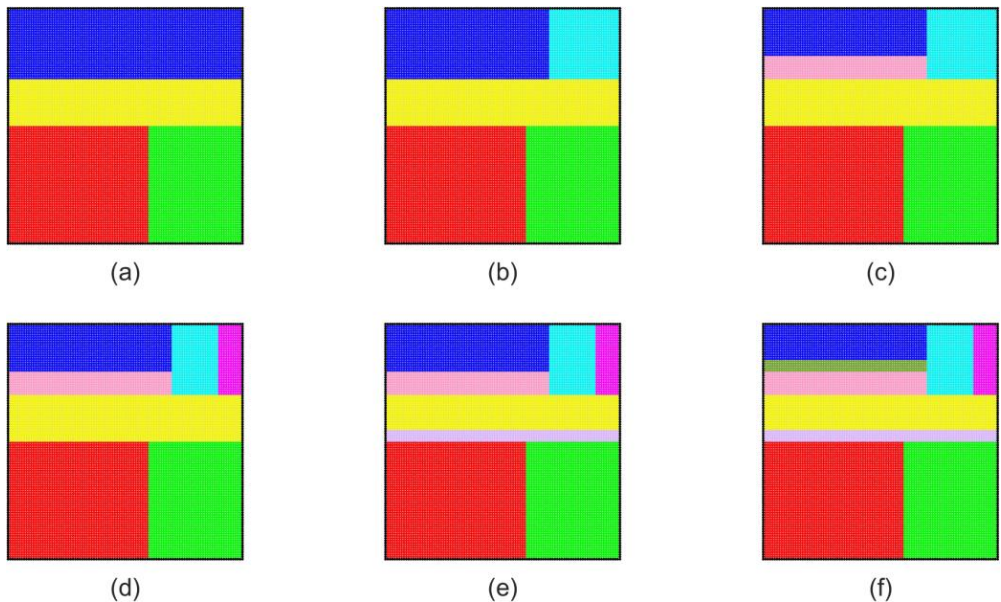

**Fig. 5.** Sample optimal solutions for different sets of rectangles used in the first study with restrained rotations. When packed optimally the objects perfectly fill the domain and their number in the sets increases from (a) to (f) as one rectangle is split into two new ones at each step.

Table 2 shows the success rates in packing all objects of each set for different optimization strategies and objective functions. For each case, the average results for 100 computations are presented. In the concurrent optimization, the effectiveness of maximizing total ones ($f_{T1}$) decreased with the increasing number of objects. Maximizing adjacent ones ($f_{A1}$) showed a similar trend but gave slightly better results for the higher number of objects. No collision avoidance constraints are considered in these studies since the higher values of the objectives also favor reducing the overlaps.

For the sequential optimization, the optimal solutions are found for all trials for the sets with 4 and 5 objects. For the analyzed sets, sorting by diagonal length generally performed better compared to area sorting, which was more slightly more effective for only intermediate numbers (6, 7) of objects. For the length sorting, using the distance to the bottom-left corner ($f_{DBL}$) as the objective function provided the best results. However, maximizing adjacent ones also worked fairly well providing a 75% success rate for 9 nine objects.

**Table 2** Success percentages in packing all objects of each set given in Table 1 for different optimization strategies and objective functions.

| Optimization strategy | | Objective function | Number of objects | | | | | |
|---|---|---|---|---|---|---|---|---|
| | | | 4 | 5 | 6 | 7 | 8 | 9 |
| Concurrent | | $f_{T1}$ | 100% | 84% | 51% | 42% | 23% | 5% |
| | | $f_{A1}$ | 100% | 78% | 42% | 46% | 27% | 9% |
| Ordered | Diagonal length ordering | $f_{DC}^r$ | 100% | 100% | 41% | 49% | 45% | 20% |
| | | $f_{DBL}^r$ | 100% | 100% | 99% | 99% | 100% | 94% |
| | | $f_{A1}$ | 100% | 100% | 81% | 81% | 73% | 75% |
| | Area ordering | $f_{DC}^r$ | 100% | 100% | 100% | 51% | 6% | 7% |
| | | $f_{DBL}^r$ | 100% | 100% | 100% | 100% | 13% | 24% |
| | | $f_{A1}$ | 100% | 100% | 98% | 100% | 9% | 6% |

For the second study, another set of objects with the quantities and dimensions specified in Table 3 is used. This set perfectly occupies the domain for optimal packing, which requires placement of the objects according to a regular pattern. For concurrent optimization, the optimal packing could not be found using $f_{T1}$ as the performance metric, where $f_{A1}$ still succeeded 6 times out of 100 trials. For sequential placement, $f_{A1}$ delivered the best results by far with its 40% success rate, where $f_{DC}$ and $f_{DBL}$ succeeded in 7% and none of the trials, respectively. The effectiveness of $f_{A1}$ originates from its inclination towards maximizing the line of contact, which in turn aids the formation of the required pattern. In Fig. 6, (a) one optimal solution for the second study and (b) positioning preferences of $f_{DC}$ and $f_{A1}$ during a sample run are demonstrated.

**Table 3** The quantities and dimensions of the rectangles used in the second study with restrained rotations.

| Quantity | Height | Width |
|---|---|---|
| 2 | 80 | 20 |
| 2 | 20 | 80 |
| 2 | 40 | 20 |
| 2 | 20 | 40 |
| 1 | 20 | 20 |

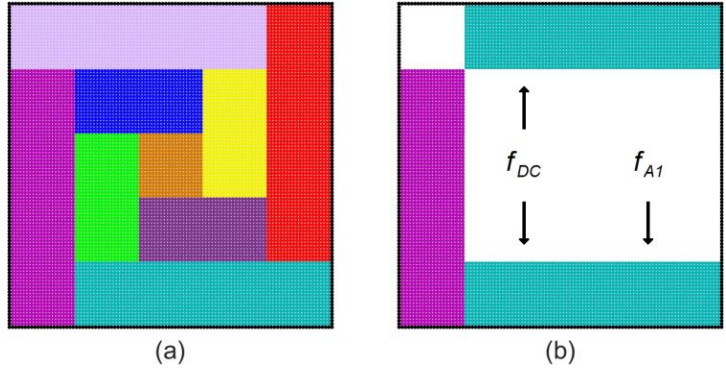

**Fig. 6.** (a) One optimal solution for the objects used in the second study with restrained rotations. The objects perfectly fill the domain for optimal packing. (b) Demonstration of two possible positions for the placement of the second object during a sample run. For $f_{DC}$ metric both positions are equally preferable whereas the bottom location is more preferable for $f_{A1}$.

*4.2. Rotatable objects*

Objects with large aspect ratios may require alteration of their orientation to be fit into the available space. New sets of objects are generated to test the developed methodologies on the packing problems with rotational degrees of freedom. The orientation angle is restricted to be a multiple of 45° to cover diagonal placement possibilities while keeping the problem size minimal.

In the first investigation, a (45 × 45) square and four smaller circles of 39 units diameter are packed into the square domain. Figure 7 shows the optimal packing configuration for these objects. Despite the relatively simple characteristics of the problem, the optimal solution cannot be attained with the ordered placement based on decreasing size. The reason is that all three performance metrics ($f_{DC}$, $f_{DBL}$, and $f_{A1}$) tend to position the large square on a corner leaving room for only three of the circles. Only using $f_{DC}$ with the small-to-big ordering leads to the optimal solution indicating that packing smaller objects first can be advantageous in certain problems. While the sequential placement performed ineffectively for this problem, concurrent placement with active constraints provided success rates of 5% and 7% for $f_{T1}$ and $f_{A1}$ metrics, respectively. Therefore, if the problem size permits its use, the concurrent optimization strategy can offer versatility in tackling different problems. As another investigation, the concurrent optimization cases are repeated by removing collision and containment constraints. In the absence of constraints, $f_{T1}$ and $f_{A1}$ still yielded 5% and 4% success rates, respectively. These results demonstrate that these metrics intrinsically facilitate the containment of rotatable objects in addition to collision avoidance.

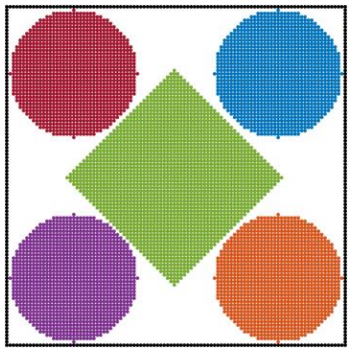

**Fig. 7.** The optimal solution for the objects used in the first study involving rotational degrees of freedom.

The second case study concerns packing three rectangles and six circles into a square domain. The quantities and dimensions of the objects used in this study are given in Table 4. With the optimal packing, all objects in the set can be placed into the domain providing a filling ratio of 74.4% as shown in Fig. 8.

The concurrent optimization strategy is not effective for packing this set due to the increased number of objects and extra rotational degrees of freedom. For this problem, the filling ratios are compared for $f_{DC}$, $f_{DBL}$, and $f_{A1}$ metrics while collision and containment constraints are held active. When the objects are ordered in decreasing length (diagonal length for the rectangles and diameter for the circles) the average filling ratios over 100 trials were found as 56.7% and 67.7% using $f_{DC}$ and $f_{DBL}$, respectively. For this ordering, the best results were achieved with $f_{A1}$, which yielded an average filling ratio of 70.8%. In this case, $f_{A1}$ outperforms the other metrics since it prefers positioning the smaller rectangles right next to the large rectangle, which is diagonally packed in the first step. Such placement leaves sufficient room at the corners where the large circles can be packed. When area-wise sorting was used, $f_{DC}$, $f_{DBL}$, and $f_{A1}$ provided 66.3%, 60.4%, and 56.5% mean filling ratios, respectively. Hence, the relative effectiveness of the utilized metrics can be significantly influenced by the ordering strategy.

**Table 4** Quantities and dimensions of the rectangular and circular objects used in the second study involving rotational degrees of freedom.

| Rectangle | | | Circle | |
|---|---|---|---|---|
| Quantity | Height | Width | Quantity | Diameter |
| 1 | 10 | 120 | 2 | 39 |
| 2 | 10 | 90 | 4 | 19 |

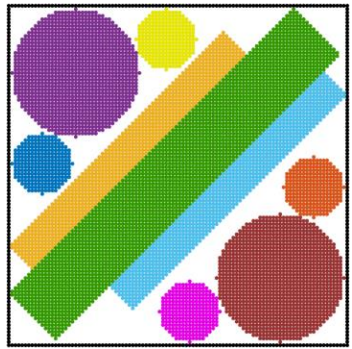

**Fig. 8.** A sample optimal solution for the objects used in the second study involving rotational degrees of freedom. The filling ratio for this packing configuration is 74.4%.

In the last study, the number of objects is increased to create a more difficult problem, where the quantities and dimensions of the packed objects are specified in Table 5. One possible optimal packing configuration for these objects is depicted in Fig. 9, where the filling ratio is 92.3%. Note that, length-wise and area-wise ordering strategies are identical for this set.

For this problem, $f_{DC}$, $f_{DBL}$, and $f_{A1}$ provided average filling ratios of 87.2%, 84.3%, and 82.2%, respectively. In this case, slight superiority of $f_{DC}$ is due to its preference of occupying the corners with the largest objects, which allows the intermediate rectangle to be diagonally placed into space in the middle. The configuration with the highest filling ratio (89.8%) is also achieved by $f_{DC}$. The $f_{DBL}$ metric provided the second-highest mean ratio and showed almost no variability over the runs. Although $f_{A1}$ gave the lowest mean filling ratio, it showed greater variability compared to $f_{DBL}$ and provided a maximum value of 88.2%, which is higher than the maximum ratio obtained by $f_{DBL}$.

**Table 5** Quantities and dimensions of the rectangular and circular objects used in the third study involving rotational degrees of freedom.

| Rectangle | | | Circle | |
|---|---|---|---|---|
| Quantity | Height | Width | Quantity | Diameter |
| 4 | 40 | 40 | 2 | 19 |
| 1 | 28 | 28 | 4 | 11 |
| 4 | 20 | 15 | | |

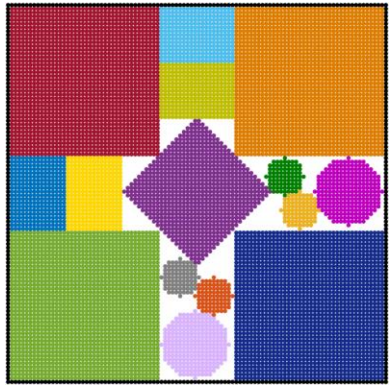

**Fig. 9.** A sample optimal solution for the objects used in the third study involving rotational degrees of freedom. The filling ratio for this packing configuration is 92.3%.

## 5. Conclusions

This study presents the new Concurrent or Ordered Matrix-based Packing Arrangement Computation Technique (COMPACT). The proposed method relies on approximating the object shapes by matrix-based (raster) representations. Such an approach provides versatility for packing various shapes and allows the use of new performance metrics.

In the proposed framework, the raster representations are obtained through loop-free operations that improve efficiency. The objects can be rotated by arbitrary angles, different from the right-angled rotation restrictions imposed in many existing packing optimization studies based on raster methods. Besides, a novel performance metric is introduced, which facilitates efficient filling of the available space by maximizing the overall contact within the domain. Moreover, exploitation of the objective functions for discarding overlap and overflow constraints is explored to enable the use of unconstrained optimization methods.

As case studies, the developed technique is utilized to pack multiple rectangular and circular objects in square bins. The suitability of concurrent and ordered placement strategies for solving different problems is investigated. In addition, the effectiveness of the utilized performance metrics is analyzed for various scenarios. The results show that the proposed technique performs effectively in obtaining plausible packing configurations. Furthermore, the case studies provide guidelines for the selection of suitable packing strategies, performance metrics, and constraints according to the problem type.

The developed techniques can certainly be extended for three-dimensional objects in future studies. It can also be used for packing problems involving various other objects and domain geometries. In particular, the new performance metric relying on contact maximization offers great potential for packing non-convex polygons. Although the genetic algorithms perform well with the described formulations, the implementation of other nonlinear optimization methods into the developed framework is also possible.

**References**

Bouzid, M. C., & Salhi, S. (2020). Packing Rectangles into a Fixed Size Circular Container: Constructive and Metaheuristic Search Approaches. *European Journal of Operational Research, 285*, 865–883.

Chen, M., & Huang, W. (2007). A two-level search algorithm for 2D rectangular packing problem. *Computers & Industrial Engineering, 53*, 123–136.

Chen, M., Tang, X., Song, T., Zeng, Z., Peng, X., & Liu, S. (2018). Greedy heuristic algorithm for packing equal circles into a circular container. *Computers & Industrial Engineering, 119*, 114–120.

Chen, M., Wu, C., Tang, X., Peng, X., Zeng, Z., & Liu, S. (2019). An efficient deterministic heuristic algorithm for the rectangular packing problem. *Computers & Industrial Engineering*, 106097.

Del Valle, A. M., de Queiroz, T. A., Miyazawa, F. K., & Xavier, E. C. (2012). Heuristics for two-dimensional knapsack and cutting stock problems with items of irregular shape. *Expert Systems with Applications, 39*, 12589–12598.

Hopper, E., & Turton, B. (1999). A Genetic Algorithm for a 2D Industrial Packing Problem. *Computers & Industrial Engineering, 37*, 375–378.

Hopper, E., & Turton, B. C. H. (2001). An empirical investigation of meta-heuristic and heuristic algorithms for a 2D packing problem. *European Journal of Operational Research, 128*, 34–57.

Jakobs, S. (1996). Theory and Methodology On genetic algorithms for the packing of polygons. *European Journal of Operational Research, 88*, 165–181.

Leao, A. A. S., Toledo, F. M. B., Oliveira, J. F., Carravilla, M. A., & Alvarez-Valdés, R. (2020). Irregular packing problems: A review of mathematical models. *European Journal of Operational Research, 282*, 803–822.

Leung, S. C. H., Zhang, D., Zhou, C., & Wu, T. (2012). A hybrid simulated annealing metaheuristic algorithm for the two-dimensional knapsack packing problem. *Computers & Operations Research, 39*, 64–73.

Martinez-Sykora, A., Alvarez-Valdes, R., Bennell, J. A., Ruiz, R., & Tamarit, J. M. (2017). Matheuristics for the irregular bin packing problem with free rotations. *European Journal of Operational Research, 258*, 440–455.

Martins, T. C., & Tsuzuki, M. S. G. (2010). Simulated annealing applied to the irregular rotational placement of shapes over containers with fixed dimensions. *Expert Systems with Applications, 37*, 1955–1972.

Mundim, L. R., Andretta, M., & de Queiroz, T. A. (2017). A biased random key genetic algorithm for open dimension nesting problems using no-fit raster. *Expert Systems with Applications, 15*, 358–371.

Sato, A. K., Martins, T. C., Gomes, A. M., & Tsuzuki, M. S. G. (2019). Raster penetration map applied to the irregular packing problem. *European Journal of Operational Research, 279*, 657–671.

Toledo F. M. B., Carravilla, M. A., Ribeiro, C., Oliveira, J. F., & Gomes, A. M. (2013). The Dotted-Board Model: A new MIP model for nesting irregular shapes. *International Journal of Production Economics, 145*, 478–487.

Zhang, W., & Zhang, Q. (2009). Finite-circle method for component approximation and packing design optimization. *Engineering Optimization, 41*, 971–987.